# A Low-temperature 1H NMR Study of H$_2$O and D$_2$O Associated Competitively with Immunoglobulin G in Solution

Andrei S. Goryunov and Alex Käiväräinen

Institute of Biology, Karelian Research Centre, Russian Academy of Sciences

Pushkinskaya 11, 185610 Petrozavodsk,

Karelia, Russia

E-mail: goryunov@krc.karelia.ru

An approach has been proposed to characterize the competitive association of D$_2$O and to study the heterogeneity of hydration water adsorbed by the protein, immunoglobulin G, using methodology for determining of non-freezing water in mixed H$_2$O-D$_2$O protein solutions by low-temperature 1H NMR technique. Direct data on the numbers of deuteriums adsorbed by immunoglobulin G and isothermals of water (D$_2$O) sorption by the protein for solution hydration conditions were obtained. The preferential binding of D$_2$O as well as the isotopic effect of low D$_2$O concentrations was simply confirmed using this method. The shape of the isothermals, similar to that for polymolecular adsorption, demonstrates relative changes in the fractions of heavy water isotope bound to different groups of protein atoms on decreasing temperature in frozen solution. At -35 °C the major fractions attached to charged and polar atomic groups appear to be related as 2/3. The adsorption curves indicate the direct relationship of non-freezing water to interface water in protein powders.

**Introduction**

Solvent interactions are key effects in the folding of a polypeptide chain and in maintaining the stability of globular proteins.[1-3] The influence of hydration on protein structure and activity has been extensively studied by a variety of methods for a long period of time.[3-5] The hydration level is now recognized to be a crucial parameter in displaying dynamic and functional properties by proteins. The mechanisms that give rise to structural dynamics, minor conformational changes and activity at a certain degree of hydration remain obscure. Both the process and values of hydration may be affected by the distorting influence of interprotein interaction in powder. In this connection it seems expedient to consider the process of water sorption under hydration conditions that differ substantially from those in moistened powder.

The anomalous effects of heavy water ($D_2O$) on the structural and functional features of proteins, as well as isotopic effects at higher levels of biological organization up to living organisms may be an illustration of the important role of hydration.[6-9] Addition of $D_2O$ was shown to enhance protein stability and aggregation and to diminish structural dynamics. The effect of $D_2O$ on protein although analogous qualitatively, differs quantitatively from that of ordinary water. Therefore, the useful practice is to study hydration in isotope mixture to gain additional information about water-protein interactions in solution. Isotope fractioning and enriching hydration water in a heavy isotope occur in (HOH-HOD-DOD) mixture. Binding of mixed water to protein was also studied mostly in adsorption experiments on moistened powders for small globular proteins.[3] Only few direct data on the amount of $D_2O$ bound to protein have been obtained in protein crystals[10,11] and films.[12] In any case the precise mechanism of preferential heavy isotope binding to proteins is still vague. The important problem is to evaluate directly the fraction of hydrated water replaced by heavy isotope in mixed isotope solution.

In this report, we describe an approach to determine the amount of bound $D_2O$ and to distinguish fractions with different hydration patterns in protein, immunoglobulin G, where low-temperature 1H NMR-spectroscopy is used to observe subsequent isotope replacement of ordinary water of hydration shell on adding heavy water (0 - 75% $D_2O$) to protein solution within the -10 - -35°C range. We present the isothermals of deuterium adsorption to protein in frozen solution. Immunoglobulin G was chosen due to its highly grooved surface and large interdomain and intersubunit cavities, which are assumed to be hydrated in a different manner in protein solution as compared to that in moistened powder.[13-15]

**Experimental**

In this study immunoglobulin G (IgG) isolated from the adult human donor blood (fresh) by gradient elution on DEAE cellulose[16,17] was used. To separate probable admixture of cryoglobulins and aggregates centrifugation at -1°C for one hour was carried out. Protein solutions were prepared on the basis of the 0.01 M phosphate buffer (pH 7.3). The protein concentration was 20 mg/ml ($\cong$ 0.13 mM).

The amount of non-freezing water in protein solution that determines protein hydration was measured by low-temperature 1H NMR within the -10 - -35°C range using Kuntz method.[18] Absolute hydration values were calculated, using a signal of the standard (24% LiCl, 0.10% $MnCl_2$ aqueous solution), which stands non-freezing up to -50°C and has water concentration 48.5 M at -35°C.[19] Temperature dependences of the integral intensity of the resonant water protons absorption band were registered using a Tesla BS-467 (60 MHz) 1H NMR spectrometer equipped by a standard thermostatic unit with a temperature maintenance accuracy of ±1°C. Superposition of five successive spectra was used as a signal averaging procedure. The outlines of an absorption peak were determined by the line going through the middle of the noise band. The spectral record of buffer solution (as a sample) was taken as a base line. In this record no buffer signal was revealed in the temperature range under study. The samples were quickly frozen in the thermostatic unit, so that the formation of a considerable amount of protein forms, differing from those observed in solution at positive temperatures,[20] was avoided. We assumed the effect of overall rearrangement of protein surface on hydration values to be negligible. Temperature dependences of the amount of non-freezing water were registered by raising temperature from the starting point -50°C. The samples were held at -50°C for 30 min and at other temperatures (up to -10°C) for 15 min. The duration of measurements in each sample was < 3 hrs.

The number of hydration water hydrogens (Hs) replaced by deuteriums (Ds) on adding $D_2O$ into protein solution is determined by the decrease in the integral intensity of the envelope resonant non-freezing water protons line, since the corresponding hydrogen fraction no longer contributes to the 1H NMR signal. The comparison of the amount of non-freezing water hydrogens in mixtures of ordinary and heavy water was carried out at pH = $pH_{measured}$, providing the identity of the macromolecule electrostatic state.[6] Dilute heavy water solutions were prepared by addition of concentrated protein solution to buffer solutions of different heavy water concentrations. The samples need not be pre-incubated due to the high rate of $H_2O + D_2O \leftrightarrow 2HOD$ equilibration.

All chemicals were reagent grade. 99.8 atomic per cent $D_2O$ was used. The data were processed using least squares, regression analysis.

**Results and discussion**

Shown in Fig. 1 are the temperature dependences of the number of non-freezing water (Hs) per IgG molecule and corresponding hydration values at different $D_2O$ concentrations in solution. The number of Hs is used to characterize the amount of non-freezing water, since in the presence of $D_2O$ a considerable number of DOH molecules appears. The ratio of DOD, DOH and HOH molecule numbers in non-freezing water can substantially differ from that in bulk water. Therefore the number of HOH molecules cannot be considered as an adequate characteristic of the amount of non-freezing water in the presence of $D_2O$. At 0% $D_2O$ concentration the number of Hs (the upper curve) is just the double number of $H_2O$ molecules per protein. The amount of non-freezing water is 710 ± 80 $H_2O$ molecules per IgG molecule or 0.09 ± 0.01 g $H_2O$/g IgG at -35°C.

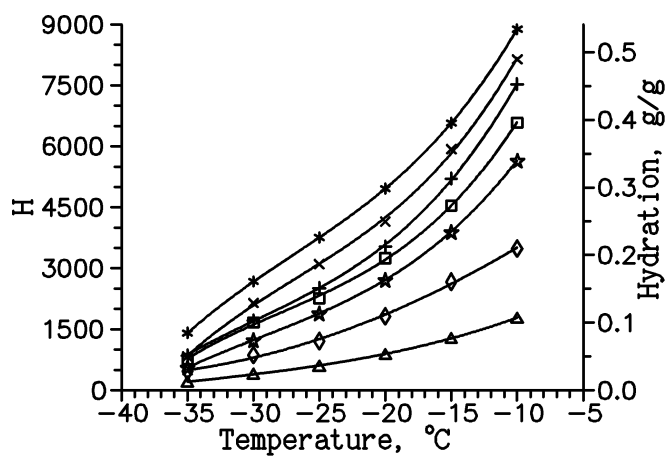

Figure 1. Temperature dependences of the number (H) of hydration hydrogens (non-freezing water hydrogens) per IgG molecule and $H_2O$ hydration of IgG (in grams of $H_2O$ per gram of protein) at different $D_2O$ concentrations: 0% (*), 5% (×), 10% (+), 20% (□), 30% (☆), 50% (◇), 75% (Δ).

The shape of the dependences (Fig. 1) reflects the increase in the amount of non-freezing water with rising temperature at all the $D_2O$ concentrations. The number of Hs decreases with the rise of $D_2O$ concentration always by the greater value than it could be expected from the equal binding of water isotopes. At -35°C the increase in $D_2O$ content from 0 to 50% in IgG solution leads to the ≅ 65% decrease in the number of Hs in non-freezing water, i.e. the preferential binding of the heavy isotope occurs, which was detected earlier by indirect methods.[6] The total HOH, HOD and DOD hydration can be higher than HOH hydration only.[21] Therefore, to characterize the preferential binding we used the number of Hs forced out (replaced) by Ds, which was calculated as a difference between the number of Hs in non-freezing water in absence of $D_2O$ ($H_{0,t}$) and that obtained at the given $D_2O$ concentration ($H_{\%,t}$) at each temperature (t) studied:

$$D_{\%,t} = H_{0,t} - H_{\%,t} \qquad (1)$$

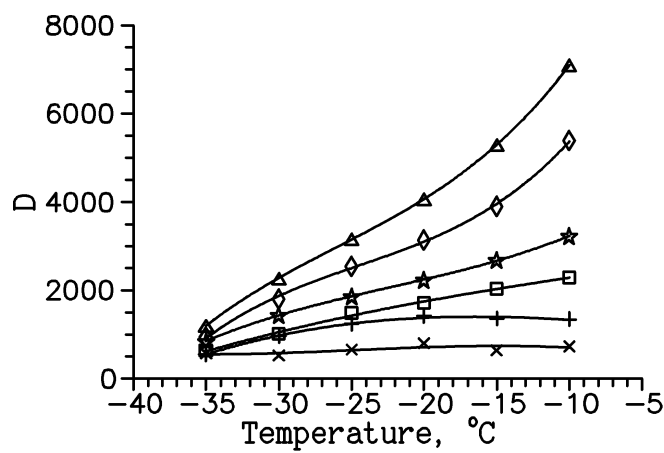

Figure 2. Temperature dependences of the number of non-freezing water hydrogens replaced by deuteriums (number of hydration deuteriums) per IgG molecule when adding different amounts of $D_2O$ to protein solution: 5% (×), 10% (+), 20% (□), 30% (☆), 50% (◇), 75% (Δ) $D_2O$.

Figs. 2 and 3 show the temperature and concentration dependences of the number of Hs replaced for Ds on adding $D_2O$ into solution. From Fig. 2 it can be seen that the number of Hs replaced by Ds at 5% $D_2O$ depends only slightly on temperature. With a rise in heavy isotope content the dependence comes to be considerably stronger. This clearly signifies that the fraction of hydrogens replaced by deuteriums at low $D_2O$ concentrations can only be affected slightly by temperature. That could be tightly bound and/or deeply buried water. On decreasing temperature the "plateau" and the "knee" are found in the dependences of the number of replaced Hs on $D_2O$ concentration with the upper limit for the "knee" at 5% $D_2O$.

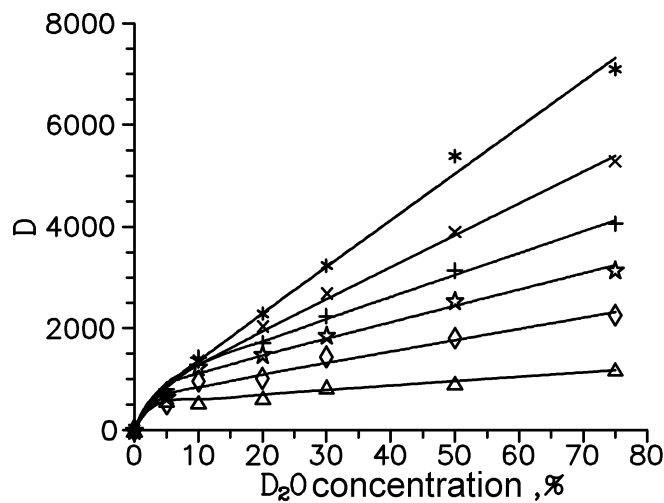

Figure 3. Dependences of the number of hydration water deuteriums per IgG molecule on $D_2O$ concentration in protein solution at -10° (*), -15° (×), -20° (+), -25° (☆), -30° (◇), -35°C (Δ).

The shape of isothermals in Fig. 3 is dominated by the effect of sequential freezing of water with lowering temperature. In order to envision an unaffected shape, we normalized the isothermals by the number of purely hydrogen hydration ($H_{0,t}$). To this end, in Fig. 4 we represent the isothermal dependences of the replaced hydrogen fraction in the total amount of non-freezing water on the D$_2$O concentration. The fraction $F_{\%,t}$ is determined as a ratio of the number of replaced Hs ($D_{\%,t}$) to the number of non-freezing water Hs in the absence of D$_2$O ($H_{o,t}$) at a given temperature:

$$F_{\%,t} = \frac{D_{\%,t}}{H_{0,t}} = \frac{H_{0,t} - H_{\%,t}}{H_{0,t}} = 1 - \frac{H_{\%,t}}{H_{0,t}} \tag{2}$$

The general shape of the isothermals represents two discernible segments. Isothermals start at 0% D$_2$O with a steep segment followed by a gently sloping segment, indicating a stepwise process of heavy isotope binding. The two stages are suggestive of different association constants of the three water forms, DOD, DOH, and HOH, whose concentration ratio change drastically on increasing deuterium content. In the initial segment (0 - 10% D$_2$O) there is a negligible DOD content and almost all the deuteriums constitute DOH form. A considerable amount ($\cong$14%) of DOD form emerges at 30% D.[6] Hence, the association constant of DOH form must be accepted higher than that of DOD to allow for the region of gentle slope. And it contradicts to the fact of preferential hydration of heavy isotope. Therefore, we suggest that the two stages result from two fractions of bound Ds and corresponding association constants.

As illustrated in Fig. 4, the isothermals of the replaced hydration hydrogen fraction on D$_2$O concentration ($F_{\%,t}$ versus deuterium concentration) are somewhat similar to adsorption isothermals for polymolecular adsorption, which were well-reported for hydration in protein powders.[3]

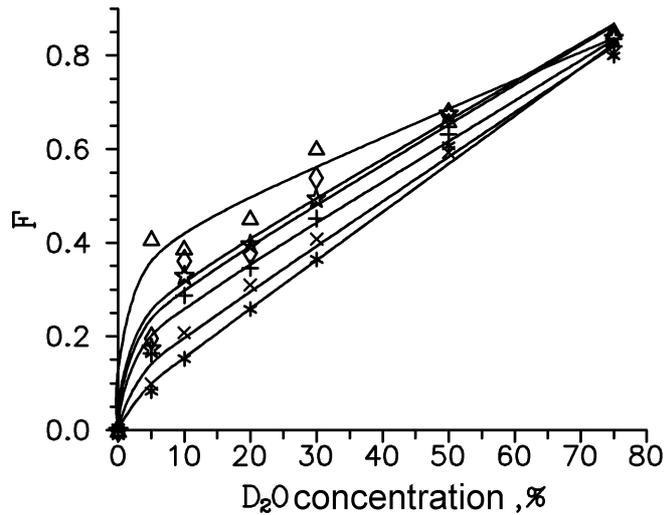

Figure 4. Dependence of the $D_2O$ fraction of unfreezing water (F) on $D_2O$ concentration in IgG solution at -10° (*), -15° (×), -20° (+), -25° (☆), -30° (◇), -35° C (Δ). Fitting curves are arranged in a descending order of the temperatures starting from the lower curve through the upper one.

The difference is that in Fig.4 the gentle segment is not followed by one more steep region at higher $D_2O$ content, which is characteristic of water adsorption in protein powders. This segment of upswing was attributed to water arranged by non-polar atomic groups. Reasons are that this analogy is not purely superficial. Firstly, the hydration Hs replacement by Ds is nothing but the adsorption of Ds forming DOD and DOH from solution due to the greater constant of heavy isotope binding by protein. Secondly, the character of interaction at DOD and DOH binding by protein in solution and in moistened powder is the same: electrostatic interactions, hydrogen bond, hydrophobic clatration, and Van der Waals forces. Thirdly, the "knee" (5% $D_2O$, Fig. 4) appears at hydration values typical for adsorption isothermals,[3] i.e. the range 0.02 - 0.03 g $H_2O$/g IgG at -35°C, 0.03 - 0.05 g/g at -30°C, 0.04 -0.07 g/g (300-540) at -25°C.

It has been generally accepted that in the "knee" range the water sorption is at charged acid and alkaline groups of side chains.[22-24] In the "plateau" range the polar groups are solvated with hydrogen bond formation, and on further hydration interaction with non-polar hydrophobic groups takes place. Water near non-polar groups (where water does not interact strongly with the protein) has been widely discussed to freeze first[25] and at -10°C water of extended surface hydrophobic segments is frozen completely.[26] Accordingly, the isothermals of Fig.4 do not display an upswing at higher $D_2O$ concentrations. Following the water of non-polar groups, the water in the vicinity of polar regions also freezes. It agrees with the fact that the "plateau" region becomes progressively less prominent as compared to the "knee". At the lower limit of the temperature range - -35°C - the largest fraction of hydration deuterium is that attached to charged side chains. It is consistent with the observation that at -35° the "knee" region comes into prominence. According to Fig. 2 this fraction is the least susceptible to temperature changes within the range studied.

Thus, the data distinguishes between two fractions of heavy isotope bound to IgG molecule. The absolute magnitudes of these fractions at -35°C can be evaluated from Fig. 3,4 as follows: no more than 570 of bound deuteriums are attached to charged sites per protein, and no less than 840 - to polar sites. The pronounced temperature dependences of deuterium fractions can result from great dynamic heterogeneity of hydration water within all the protein atom classes which follows from residence times observations:[27,28] waters characterized by longer residence times freeze at lower temperatures. Analogy of the isothermals of heavy water isotope adsorption on protein in solution to those reported for water adsorption in dry protein suggests the direct relationship of the non-freezing water to the interface water in protein powders. These results appear to clear up the conflict[3] of measurements on partially hydrated proteins with the suggestion by Kuntz[29] that the non-freezing water is dominated by contribution from water about ionic residues. The specific chemistry and geometry of internal water - protein interactions can also be a source of that contradiction, particularly for multidomain oligomeric proteins like IgG with their interdomain and intersubunit cavities. The residence times of cavity waters are even longer than that of water molecules covering charged and polar atoms.[27,28] Allowing for the "knee" to occur at about 5% $D_2O$, the anomalous isotopic effect of low $D_2O$ concentrations should be assigned to the adsorption of deuteriums at charged groups of protein atoms and other sites with longest residence times.

The further attribution of hydration water fractions in solution to certain types of binding sites in protein by means of this approach is possible by comparing proteins in ligand states that differ, for instance, in the quantity of charged groups at the surface of a globule or in the sizes of intramolecular cavities. From this point of view the appropriate object is hemoglobin in oxy, deoxy and met forms.